\documentclass[twocolumn,showpacs,prl,amsmath,amssymb,superscriptaddress]{revtex4}
\usepackage{graphicx}% Include figure files
\usepackage{dcolumn}% Align table columns on decimal point
\usepackage{bm}% bold math
\usepackage{amssymb}

\begin{document}

\title{Manipulation and Generation of Supercurrent in Out-of-Equilibrium Josephson Tunnel Nanojunctions}

\author{S. Tirelli}
\affiliation{NEST CNR-INFM and Scuola Normale Superiore, I-56126 Pisa, Italy}
\author{A. M. Savin}
\affiliation{Low Temperature Laboratory, Helsinki University of Technology, P.O. Box 3500, FIN-02015 TKK, Finland}
\author{C. Pascual Garcia}
\affiliation{NEST CNR-INFM and Scuola Normale Superiore, I-56126 Pisa, Italy}
\author{J. P. Pekola}
\affiliation{Low Temperature Laboratory, Helsinki University of Technology, P.O. Box 3500, FIN-02015 TKK, Finland}
\author{F. Beltram}
\affiliation{NEST CNR-INFM and Scuola Normale Superiore, I-56126 Pisa, Italy}
\author{F. Giazotto}
\email{giazotto@sns.it}
\affiliation{NEST CNR-INFM and Scuola Normale Superiore, I-56126 Pisa, Italy}
%\date{\today}

\begin{abstract}
We demonstrate experimentally manipulation of supercurrent in
Al-AlO$_x$-Ti Josephson tunnel junctions by injecting quasiparticles
in a Ti island from two additional tunnel-coupled Al superconducting
reservoirs. Both supercurrent enhancement and quenching with respect
to equilibrium are achieved. We demonstrate cooling of the Ti line
by quasiparticle injection from the normal state deep into the
superconducting phase. A model based on heat transport and
non-monotonic current-voltage characteristic of a Josephson junction
satisfactorily accounts for our findings.
\end{abstract}


\maketitle

Nonequilibrium dynamics in superconducting nanocircuits is currently
in the focus of an intense experimental and theoretical effort
\cite{kopnin,rmp}. In this context, the control of the Josephson
current in superconductor-normal metal-superconductor (SNS) weak
links  is receiving much attention. In these systems supercurrent is
manipulated by modifying the quasiparticle energy distribution in
the N region via current injection from external terminals
\cite{wilhelm,volkov,giazotto04}. There have been some successful
demonstrations of such out-of-equilibrium SNS junctions
\cite{baselmans,savin,morpurgo,crosser06}. On the other hand, it was
predicted \cite{giazotto05,laakso} that supercurrent can be
controlled  in all-superconducting tunnel structures as well. In
this case  quasiparticle injection
\cite{blamire,ammendola,brinkman,nevirkovets,ketterson,manninen},
leads to intriguing features peculiar to out-of-equilibrium
superconductors.

In this Letter we report on control of the
Josephson coupling in a small S island by injecting quasiparticles
from tunnel-coupled superconducting leads. Both supercurrent
enhancement and suppression with respect to equilibrium, as well as
generation at temperatures above the island critical temperature
were achieved by changing the quasiparticle injection rate. Our
findings are explained within a model relating the superconducting state of the
island to the heat flux driven through it upon injection.

\begin{figure}[t!]
\includegraphics[width=\columnwidth,clip]{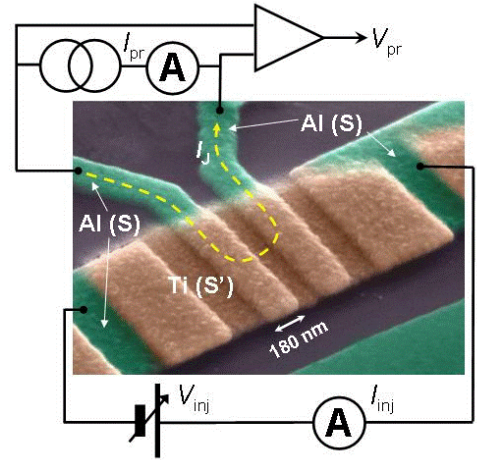}
\caption{(color online) A typical structure (Sample B) showing a
schematic of the measurement setup. In the middle, a Ti superconducting
island (S$^{'}$) is connected to four Al electrodes (S) through
tunnel junctions. $I_\text{J}$ denotes the Josephson current flowing
through the two inner Al-AlO$_x$-Ti tunnel junctions.} \label{fig1}
\end{figure}
Figure \ref{fig1} shows a scanning electron micrograph  of a typical
structure along with a scheme of the measurement setup. The core of
the sample consists of a SIS$^{'}$IS control line, i.e., a titanium
(Ti) superconducting island (S$^{'}$) symmetrically connected at its
ends via AlO$_x$ barriers (I) with normal-state resistance
$R_{\text{T}}$ each to two aluminum (Al) superconducting reservoirs
(S). Two additional Al-AlO$_x$-Ti probe junctions, with normal-state
resistance $R_{\text{J}}$ each and placed in the center of the
island, are used to measure the Josephson current ($I_{\text{J}}$)
of their nominally symmetric series connection.
%Al-AlO$_x$-Ti
%structures were also considered in Ref. \cite{manninen},  but
%operated in the context of electron refrigeration \cite{rmp}.
Our
samples were fabricated by electron beam lithography and two-angle
shadow-mask evaporation. The measurements were performed in a
dilution refrigerator at sub-kelvin temperatures measured with a
RuO$_2$ resistor calibrated against Coulomb blockade thermometer
\cite{rmp}. The experiment consists of measuring at different bath
temperatures ($T_{\text{bath}}$) the current-voltage characteristic
($I_{\text{pr}}\text{ vs }V_{\text{pr}}$) of the series connection
of the central SIS$^{'}$ Josephson junctions while imposing a fixed
voltage ($V_{\text{inj}}$) across the lateral Al reservoirs. As we
shall show, this will lead to a change in temperature of S$^{'}$
which determines the dynamics of the Josephson junctions.

\begin{figure}[t!]
\includegraphics[width=\columnwidth,clip]{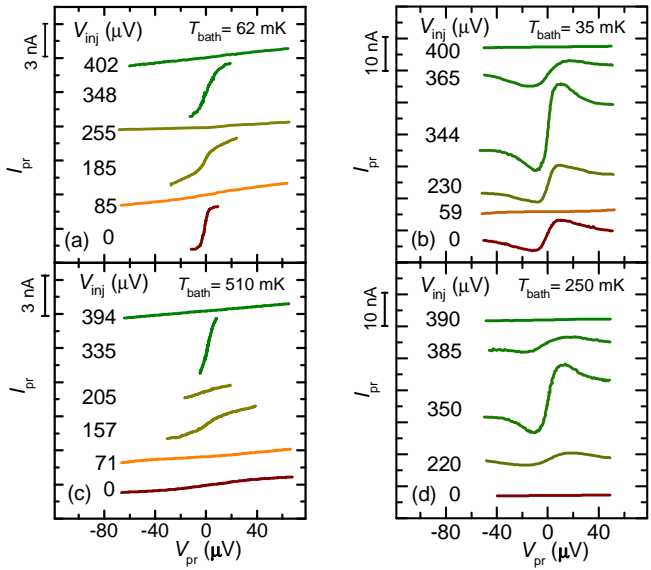}
\caption{(color online) $I_{\text{pr}}\text{ vs }V_{\text{pr}}$ for various values of $V_{\text{inj}}$:
(a) Sample A, $T_{\text{bath}}=62$ mK; (b) Sample B, $T_{\text{bath}}=35$ mK; (c) Sample A, $T_{\text{bath}}=510$ mK;
(d) Sample B, $T_{\text{bath}}=250$ mK.
The curves are vertically offset for clarity.
}
\label{fig2}
\end{figure}

Figure \ref{fig2} shows the electrical characterization of two
structures, in the following referred to as Sample A (whose
essential parameters are $R_{\text{T}}\simeq 1.43$ k$\Omega$,
$R_{\text{J}}\simeq 2.8$ k$\Omega$ and a 45-nm-thick Ti island of
area $250\times 2550$ nm$^2$), and Sample B (with
$R_{\text{T}}\simeq 710$ $\Omega$, $R_{\text{J}}\simeq 1.56$
k$\Omega$ and a 40-nm-thick Ti island of area $650\times 1500$
nm$^2$). The critical temperature ($T_{\text{c}}^{'}$) of the Ti
layer is $\sim 500$ mK for Sample A and $\sim 210$ mK for Sample B.
Panels (a) and (b) display the low-temperature $I_{\text{pr}}\text{
vs }V_{\text{pr}}$ characteristics of Sample A and B, respectively,
for several values of the injection voltage $V_{\text{inj}}$. Each
characteristic corresponds to a different $V_{\text{inj}}$, and the
curves are vertically offset for clarity. In equilibrium, at
$V_{\text{inj}}=0$, the supercurrent manifests itself as a peak
around zero bias in the current-voltage characteristic. 
As will be
explained with further details, upon increasing  the injection
voltage the supercurrent behaves non-monotonically, 
being initially
suppressed, 
then showing typically two peaks at intermediate injection voltages.
Further increase of $V_{\text{inj}}$ leads to a monotonic
supercurrent decay, and to a complete quenching for
$V_{\text{inj}}\gtrsim 400\,\mu$V. In Sample B the peak amplitude is
enhanced by almost a factor of three with respect to equilibrium.
The supercurrent response in the high-temperature regime [see panels
(c) and (d) for Sample A and B, respectively] is  different. In
particular, while the equilibrium supercurrent is already vanishing,
as $T_{\text{bath}}$ exceeds the critical temperature of the Ti
island, it is generated by increasing $V_{\text{inj}}$ at an
injection voltage which cools S$^{'}$ from the normal into the
superconducting state. This occurs thanks to hot quasiparticle
extraction provided by the Al reservoirs
\cite{rmp,giazotto05,laakso}. By increasing $V_{\text{inj}}$ even
further  leads to another maximum of supercurrent followed by full
suppression. Although somewhat different in terms of characteristic
parameters, both samples show similar behavior.
\begin{figure}[t!]
\includegraphics[width=\columnwidth,clip]{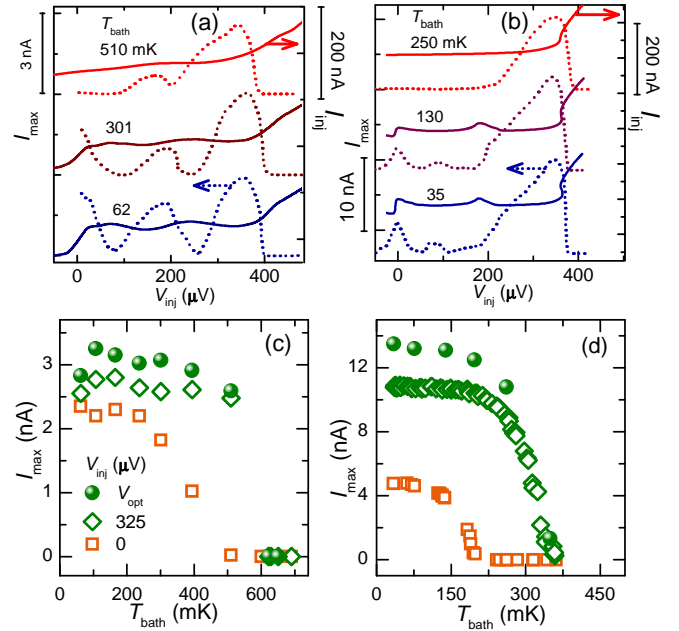}
\caption{(color online) (a) Left axis: $I_{\text{max}}$ vs
$V_{\text{inj}}$ at three different $T_{\text{bath}}$ for Sample A.
Right axis: Injector characteristics $I_{\text{inj}}\text{ vs
}V_{\text{inj}}$ at the same bath temperatures. (b) The same as in
(a) for Sample B. (c) $I_{\text{max}}$ vs $T_{\text{bath}}$ for
three different values of $V_{\text{inj}}$ for Sample A. (d) The
same as in (c) for Sample B. } \label{fig3}
\end{figure}

The full dependence of the maximum supercurrent $I_{\text{max}}$ on
$V_{\text{inj}}$ at different bath temperatures is displayed on the
left axis of Fig. \ref{fig3}(a) and (b) for Sample A and B,
respectively. $I_{\text{max}}$ is defined as the average between the
amplitudes of positive and negative  peaks of $I_{\text{pr}}$. It is
a symmetric function of $V_{\text{inj}}$ based on electron-hole
symmetry, so that just the dependence on positive $V_{\text{inj}}$
is shown. As we shall show in the following, the features present at
small $V_{\text{bias}}$ in the supercurrent response are related to
the shape of the current-voltage characteristic of the injectors
($I_{\text{inj}}$ vs $V_{\text{inj}}$), shown on the right axis of
Fig. \ref{fig3}(a) and (b) for the same $T_{\text{bath}}$. In
particular, in addition to a current enhancement around
$V_{\text{inj}}=0$ originating from Josephson coupling in the
lateral SIS$^{'}$ junctions, the curves at lower $T_{\text{bath}}$
show a marked peak centered around the middle of the characteristic
which disappears as soon as S$^{'}$ undergoes a transition into the
normal state.

Figures \ref{fig3}(c) and (d) show the $I_{\text{max}}$ vs
$T_{\text{bath}}$ characteristic for Sample A and B, respectively,
at three different values of $I_{\text{inj}}$. For
$V_{\text{inj}}=0$ (open squares) the equilibrium supercurrent
saturates at low $T_{\text{bath}}$ where it obtains values as high
as $\simeq 2.35$ nA and $\simeq 4.8$ nA for Sample A and B,
respectively, while it is gradually reduced by increasing the
temperature, being completely suppressed at $T_{\text{bath}}\simeq
500$ mK and $\simeq 210$ mK, i.e., at the critical temperature of
Sample A and B, respectively. The low-temperature supercurrent
amplitudes are suppressed in both samples by about an order of
magnitude as compared to the Ambegaokar-Baratoff theoretical
prediction \cite{ambegaokar}. This is however expected for
ultrasmall Josephson tunnel junctions influenced by environment
fluctuations \cite{steinbach,zilberman}. For a chosen injection
voltage, e.g., at $V_{\text{inj}}=325\,\mu$V, $I_{\text{max}}$
saturates at low $T_{\text{bath}}$ at $\simeq 2.55$ nA for Sample A,
and $\simeq 10.8$ nA for Sample B. The maximum supercurrent survives
under injection up to $T_{\text{bath}}\simeq 630$ mK for Sample A
and $\simeq 360$ mK for Sample B, i.e., well above the equilibrium
critical temperature. This means that we can cool the samples by
quasiparticle current from the normal into the superconducting
state. Also shown is the temperature dependence at the optimized
bias voltage ($V_{\text{opt}}$) which maximizes $I_{\text{max}}$
(solid dots).

\begin{figure}[t!]
\includegraphics[width=\columnwidth,clip]{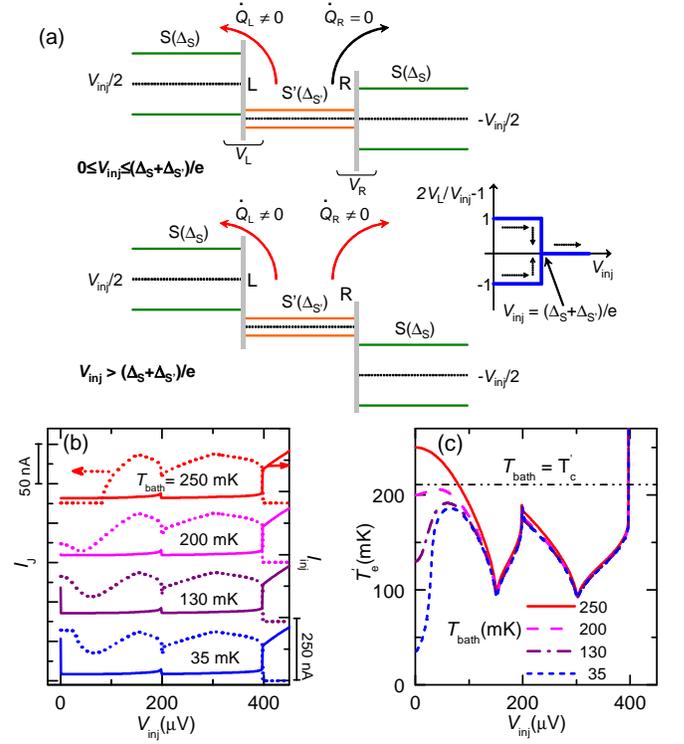}
\caption{(color online) (a) Energy-band diagram of the biased
SIS$^{'}$IS control line. Also shown on the right side is the voltage drop
$V_{\text{L}}$ across left interface. (b) Left axis: $I_{\text{J}}$
vs $V_{\text{inj}}$ calculated for a few values of
$T_{\text{bath}}$. Right axis: $I_{\text{inj}}\text{ vs
}V_{\text{inj}}$ calculated for the same $T_{\text{bath}}$. The
curves are vertically offset for clarity. (c) Calculated
$T_{\text{e}}^{'}$ vs $V_{\text{inj}}$ for the same
$T_{\text{bath}}$ as in (b). The horizontal line indicates Ti
critical temperature of Sample B.}
\label{fig4}
\end{figure}

Our observations of non-monotonic dependence of the probe
supercurrent on bias voltage and of a peak in the current in the
middle of the superconducting gap can both be explained
qualitatively within a very simple model. The key observation is
that the bias voltage dependence of the current of a single injector
junction is non-monotonic because it can be carried by Cooper pairs
(supercurrent around zero voltage) and by quasiparticles (near and
above the gap voltage). Then, as a function of bias voltage
$V_{\text{inj}}$ across the two injecting junctions, the evolution
of the voltage across each individual junction is as follows [see
the energy-band diagrams in Fig. \ref{fig4}(a)]. At around zero
bias, both junctions carry supercurrent, seen as a peak in the
current-voltage characteristic. Thereabove one of the junctions,
i.e., the one with smaller critical current [for instance, left (L)
injector in Fig. \ref{fig4}(a)], switches into the quasiparticle
branch: the total voltage then equals that across this "weaker"
junction, while the other one remains in the approximately zero
voltage supercurrent branch. In this situation, when the voltage is
approximately $(\Delta_{\rm S}-\Delta_{\rm S'})/e$
\cite{rmp,giazotto05,laakso}, there is increase of Josephson
critical current of the probe junctions, thanks to enhanced cooling
power ($\dot{Q}_{\text{L}}$) due to quasiparticle current in the L
junction, i.e., $\dot{Q}_{\text{L}}\neq 0$. Here
$\Delta_{\text{S,S}^{'}}$ is the BCS energy gap in S (S$^{'}$). In
the middle of the gap region at $(\Delta_{\rm S}+\Delta_{\rm
S'})/e$, one of the junctions reaches the steep onset of
quasiparticle current leading to a peak in control current. Above
this bias, also the second junction [i.e., right (R) injector in
Fig. \ref{fig4}(a)] switches into quasiparticle branch providing
finite cooling power, i.e., $\dot{Q}_{\text{R}}\neq0$. Now the
voltage is divided approximately equally across the two junctions,
and at intermediate voltages above $(\Delta_{\rm S}+\Delta_{\rm
S'})/e$ cooling power is small until it maximizes at $2(\Delta_{\rm
S}-\Delta_{\rm S'})/e$ \cite{rmp,giazotto05,laakso} resulting in
another maximum in probe supercurrent. The final increase of current
$I_{\text{inj}}$ in the control junctions occurs at $2(\Delta_{\rm
S}+\Delta_{\rm S'})/e$, where both junctions have an approximately
equal voltage corresponding to the onset of quasiparticle current:
this results in large current, heating of the S$^{'}$ island, and
subsequent quench of the probe supercurrent.

A more quantitative analysis can be carried out as follows. The
total electric current flowing through left and right injectors can
be written as
$I_{\text{inj}}^{\text{L,R}}=I_{\text{J}}^{\text{L,R}}+I_{\text{qp}}^{\text{L,R}}$,
where $I_{\text{J}}^{\text{L,R}}\neq 0$ for $V_{\text{L,R}}=0$ is
the Ambegaokar-Baratoff critical current of the injectors
\cite{ambegaokar}, $V_{\text{L,R}}$ is the voltage drop across L(R)
interface [see Fig. \ref{fig4}(a)], while
$I_{\text{qp}}^{\text{L,R}}=\pm\frac{1}{eR_{\text{T}}}\int
d\epsilon\mathcal{N}_{\text{S}}(\tilde{\epsilon}_{\text{L,R}})
\mathcal{N}_{\text{S}^{'}}(\bar{\epsilon})[f_0(\tilde{\epsilon}_{\text{L,R}},
T_{\text{bath}})-f_0(\bar{\epsilon},T_{\text{e}}^{'})]$ is the
quasiparticle current. Here,
$\tilde{\epsilon}_{\text{L,R}}=\epsilon\mp eV_{\text{inj}}/2$,
$\bar{\epsilon}=\epsilon-e(V_{\text{inj}}/2-V_{\text{L}})$,
$f_0(\epsilon,T)$ is the Fermi-Dirac function at temperature $T$,
and $\mathcal{N}_{\text{S,S}^{'}}(\epsilon)$ is the smeared density
of states of S(S$^{'}$). In particular we set
$\mathcal{N}_{\text{S,S}^{'}}(\epsilon)=|\text{Re}[(\epsilon+i\Gamma_{\text{S,S}^{'}})/\sqrt{(\epsilon+i\Gamma_{\text{S,S}^{'}})^2-\Delta^2_{\text{S,S}^{'}}}]|$,
where $\Gamma_{\text{S,S}^{'}}$ accounts for quasiparticle states
within the gap in S(S$^{'}$) \cite{rmp}.

The voltage drop across L(R) interface resulting from biasing with
$V_{\text{inj}}$ follows from the conservation of the total current,
i.e., $I_{\text{inj}}^{\text{L}}=I_{\text{inj}}^{\text{R}}$ with
$V_{\text{L}}+V_{\text{R}}=V_{\text{inj}}$. The solution for
$V_{\text{L}}$ [shown on the right side of Fig. \ref{fig4}(a)] is
$2V_{\text{L}}/V_{\text{inj}}-1=\pm 1$ for $0\leq V_{\text{inj}}\leq
(\Delta_{\text{S}}+\Delta_{\text{S}^{'}})/e$, and
$V_{\text{L}}=V_{\text{inj}}/2$ for
$V_{\text{inj}}>(\Delta_{\text{S}}+\Delta_{\text{S}^{'}})/e$,
meaning that only one junction is initially in the dissipative
regime [L (R) junction in the upper (lower) branch]. The threshold
for equal voltage division is $V_{\text{inj}}\simeq
(\Delta_{\text{S}}+\Delta_{\text{S}^{'}})/e$ and it depends only
marginally on the asymmetry between the two injector junctions.

The supercurrent of the probe junctions depends on the quasiparticle
distribution in S$^{'}$ under voltage biasing
\cite{giazotto05,laakso}. Strong electron-electron interaction
drives the electron system in S$^{'}$ into local thermal
(quasi)equilibrium described by a Fermi-Dirac function at an
electron temperature $T_{\text{e}}^{'}$ which may differ from
$T_{\text{bath}}$ \cite{rmp}. The maximum Josephson current flowing
through the central SIS$^{'}$ junctions is given by
\cite{giazotto05,laakso}
\begin{eqnarray} 
\label{josephson}
I_{\text{J}}=\frac{1}{2eR_{\text{J}}}\big{|} \int
d\epsilon\{[1-2f_0(\epsilon,T_{\text{e}}^{'})]\text{Re}[\mathcal{F}_{\text{S}^{'}}(\epsilon)]\text{Im}[\mathcal{F}_{\text{S}}(\epsilon)]\\\nonumber
+[1-2f_0(\epsilon,T_{\text{bath}})]\text{Re}[\mathcal{F}_{\text{S}}(\epsilon)]\text{Im}[\mathcal{F}_{\text{S}^{'}}(\epsilon)]\}\big{|},
\end{eqnarray}
where
$\mathcal{F}_{\text{S,S}^{'}}(\epsilon)=\Delta_{\text{S,S}^{'}}/\sqrt{(\epsilon
+i\Gamma_{\text{S,S}^{'}})^2-\Delta^2_{\text{S,S}^{'}}}$. In the
above expressions we set
$\Delta_{\text{S}}=\Delta_{\text{S}}(T_{\text{bath}})$ and
$\Delta_{\text{S}^{'}}=\Delta_{\text{S}^{'}}(T_{\text{e}}^{'})$. 
Equation (1)  shows that $I_{\text{J}}$ is
controlled by $T_{\text{e}}^{'}$ once $T_{\text{bath}}$ is fixed.
Under bias voltage $V_{\text{inj}}$ the heat current
($\dot{Q}_{\text{L,R}}$) flowing from S$^{'}$ to S through L or R
interface is given by \cite{manninen,frank}
\begin{eqnarray}
\dot{Q}_{\text{L,R}}=\frac{1}{e^2R_{\text{T}}}\int
d\epsilon\bar{\epsilon}\mathcal{N}_{\text{S}}(\tilde{\epsilon}_{\text{L,R}})\mathcal{N}_{\text{S}^{'}}(\bar{\epsilon})\\\nonumber
\times [f_0(\bar{\epsilon},T_{\text{e}}^{'})-f_0(\tilde{\epsilon}_{\text{L,R}},T_{\text{bath}})].
\end{eqnarray}
$T_{\text{e}}^{'}$ is then determined by solving the
energy-balance equation
$\dot{Q}_{\text{L}}(V_{\text{inj}},T_{\text{bath}},T_{\text{e}}^{'})
+\dot{Q}_{\text{R}}(V_{\text{inj}},T_{\text{bath}},T_{\text{e}}^{'})=0$.
We neglect the electron-phonon interaction contribution in the
energy-balance equation which would lead to small corrections only.
The probe supercurrent is then determined by the electron
temperature $T_{\text{e}}^{'}$ established in S$^{'}$ by biasing the
control line.

For comparison with the experiment we chose the given parameters of
Sample B, $T_{\text{c}}=1.2$ K and depairing parameter
$\Gamma_{\text{S(S}^{'})}=5\times 10^{-3}\Delta_{\text{S(S}^{'})}$.
The injector current-voltage characteristics calculated at different
$T_{\text{bath}}$ are displayed on the right axis of Fig.
\ref{fig4}(b). In addition to Josephson coupling vanishing at
$T_{\text{bath}}\geq T_{\text{c}}^{'}$, the current shows a peak
centered in the middle of the characteristic, as observed in the
experiment [see right axis of Figs. \ref{fig3}(a-b)]. The
$I_{\text{J}}$ vs $V_{\text{inj}}$ characteristics are displayed on
the left axis of Fig. \ref{fig4}(b) for the same $T_{\text{bath}}$
values. The supercurrent curves of Fig. \ref{fig3}(a-b) resemble
those of the model presented in Fig. \ref{fig4}(a), apart from
details that we attribute to the oversimplified thermal model.

Figure \ref{fig4}(c) shows the electron temperature
$T_{\text{e}}^{'}$ calculated from the energy-balance equation for
the corresponding bath temperatures. For $T_{\text{bath}}\leq 200$
mK the electron gas is initially heated, inducing supercurrent
suppression at small bias voltages. Such heating stems from subgap
current in a tunnel junction \cite{rmp,pekola04,rajauria}. By increasing
$V_{\text{inj}}$ further, the electron temperature starts to
decrease, thanks to quasiparticle cooling \cite{rmp,frank} provided
by the larger gap superconductor (S), and is minimized at
$V_{\text{inj}}\simeq 150\,\mu$V. Further increase of bias voltage
leads again initially to heating, then cooling, and eventually
heating above $T_{\text{c}}^{'}$ for large $V_{\text{inj}}$. At the
bath temperature of $250$ mK, $T_{\text{e}}^{'}$ starts to decrease
monotonically, initially driving S$^{'}$ into the superconducting
state, and showing the same behavior as at lower $T_{\text{bath}}$.

In conclusion, control of Josephson current as well as its
generation at bath temperatures above the critical one were achieved
by varying quasiparticle injection into a small superconducting
island. Our results are successfully described within a model
relating the superconducting state of the island to the heat flux
originating from quasiparticle injection.
From the practical point of view, our experiment demonstrates that
quasiparticle injection can cool a metal wire from its normal state
deep into the superconducting phase.

We acknowledge financial support from the EU Large Scale
Installation Program ULTI-3 and from the NanoSciERA "NanoFridge"
project.

\end{document}